\documentclass[twocolumn,epjc3]{svjour3}

\journalname{Eur. Phys. J. C}

\usepackage{graphicx}
\usepackage{relsize}
\usepackage[a4paper,margin=2cm]{geometry}
\usepackage{amssymb}
\usepackage{amsmath}
\usepackage{booktabs}
\usepackage[switch, displaymath, mathlines, columnwise]{lineno}
\newcommand{\GG}{{\gamma\gamma}}
\begin{document}
\title{
On semi-exclusive measurement of $\GG\to\GG$ scattering%
}

\author{Rafa\l{} Staszewski\thanksref{e,addr1}
\and
Janusz J. Chwastowski\thanksref{addr1}
}

\thankstext{e}{e-mail: rafal.staszewski@ifj.edu.pl}

\institute{
Institute of Nuclear Physics Polish Academy of Sciences, ul. Radzikowskiego 152, 31-342 Krak\'ow, Poland\\[7mm]
\label{addr1}
}


\maketitle
\begin{abstract}
  The two-photon production of photon pairs, \textit{i.e.} the $\GG\to\GG$ process, is studied.
  Different production modes regarding the elastic or inelastic coupling of the intermediate-state photons to the protons are considered.
  The semi-exclusive measurement, where one intact proton is registered by a dedicated forward proton detector, is discussed.
  As an example, the signal and background simulations are performed for the $\GG\to\GG$ process mediated by the hypothetical 750 GeV resonance.
\end{abstract}

\section{Introduction}

The recent indication of the possible existence of a new resonance of mass around 750 GeV decaying into two photons \cite{Aaboud:2016tru,CMS:2015dxe}, reported by the ATLAS and CMS experiments at the Large Hadron Collider, has been enthusiastically received by the community and triggered a great interest in its possible origin (see \textit{e.g.} \cite{Franceschini:2015kwy,Ellis:2015oso,Lebiedowicz:2016lmn}).
One of the possibilities is based on a simple observation that a particle decaying into two photons can be also produced in the photon--photon fusion.

Since a photon can couple to the proton elastically, without breaking it apart, it is possible to consider the following process ($R$ is the hypothetical 750~GeV $\GG$ resonance):
\begin{equation}
  \nonumber
  pp \to p + \gamma^\ast \gamma^\ast+ p \to p+R+p \to p+\GG+ p.
  \label{}
\end{equation}

Such a process provides an interesting possibility of the exclusive measurement,
where all final-state particles are registered.
This can be achieved with dedicated forward proton detectors,
which can register protons scattered at very small angles into the accelerator beam pipe.
The ATLAS and CMS/TOTEM experiments at the LHC are already equipped with such systems: the ATLAS Forward Proton (AFP) detectors \cite{bib:afp} and the CMS-TOTEM Proton Precision Spectrometer (CT-PPS) \cite{bib:pps}.

If the existence of the 750 GeV resonance is confirmed, 
an exclusive measurement would provide additional information about its nature.
In particular, such a measurement would ensure the two-photon production mechanism of the resonance.
This is because the gluon-mediated exclusive processes at high masses are suppressed with respect to the photon-mediated ones~\cite{Khoze:2001xm}.

The exclusive measurement with two photons in the final state was already discussed in the context of general searches for new physics via anomalous $\gamma\gamma\gamma\gamma$ couplings \cite{Fichet:2014uka}
and also in the context of the hypothetical $\gamma\gamma$ resonance \cite{Fichet:2016pvq}.

In this letter, for the first time, a possibility of a semi-exclusive measurement of the light-by-light scattering is discussed.
A signal and background simulation is performed and the effects of (semi-)exclusivity selection are studied.
The analysis is performed for the example case of the 750 GeV resonance.
However, the presented method is more general and can be used for studies of $\GG\to\GG$ processes in a wide range of invariant masses.

\section{Two-photon processes}

The state-of-the-art calculation of the cross sections for the two-photon production can be found in \cite{Harland-Lang:2016qjy}.
The authors report 15\% -- 20\% uncertainties of their results.
The present study does not attempt to approach such a high precision;
the goal is rather to discuss the measurement feasibility with a relatively simple and straightforward model.
Nevertheless, the following calculations are expected to provide characteristics of the events that are sufficiently close to the reality.

The model is based on the equivalent photon approximation, 
in which the cross section for a given two-photon process factorises into 
the cross section for the photon--photon interaction and the photon--photon luminosity:
\begin{equation}
  \nonumber
  \text{d}\sigma 
  = 
  \sigma(\GG\to X) 
  \,
  \text{d}\mathcal{L}_\GG.
\end{equation}
The photon--photon luminosity is a product of the fluxes of both interacting photons:
\begin{equation}
  \nonumber  
  \text{d}\mathcal{L}_\GG = \text{d}\Phi_1\,\text{d}\Phi_2.
\end{equation}

For the following study it is natural to single out two possible types of photon emission.
First, the photon can be emitted coherently by the proton, which then remains intact after the interaction.
Second, the photon can be emitted inelastically,
\textit{e.g.} when it couples directly to a single quark,
which leads to the dissociation of the proton.

In this work the elastic photon flux is taken from the approximate parameterisation \cite{Drees:1988pp}:
\begin{multline}
  \nonumber
  \Phi(x) = 
  \frac{\alpha}{2\pi x}
  \left[ 1 + (1-x)^2 \right]
  \\
  \left[ \log A - \frac{11}{6} + \frac{3}{A} - \frac{3}{2A^2} + \frac{1}{3A^2} \right], 
\end{multline}
where $x$ is the proton momentum fraction carried by the photon, 
$A = 1 + (0.71 \text{GeV}^2)/Q^2_\text{min}$
and
\begin{multline}
  \nonumber
  Q^2_\text{min} = 
  -2m_p^2 + \frac{1}{2s} 
  \bigg[
  (s+m_p^2)(s - xs + m_p^2)
  \\
  - (s-m_p^2)\sqrt{(s-xs-m_p^2) - 4m_p^2xs}
  \bigg]
  ,
\end{multline}
with $m_p$ being the proton mass and $s$ the centre-of-mass energy squared.
The inclusive photon flux is taken from the MRST2004 QED parton distribution set \cite{Martin:2004dh,Buckley:2014ana}.
The inelastic flux is taken as the difference between the inclusive and the elastic flux.

\begin{figure}[b]
  \includegraphics[width=\linewidth, page=1]{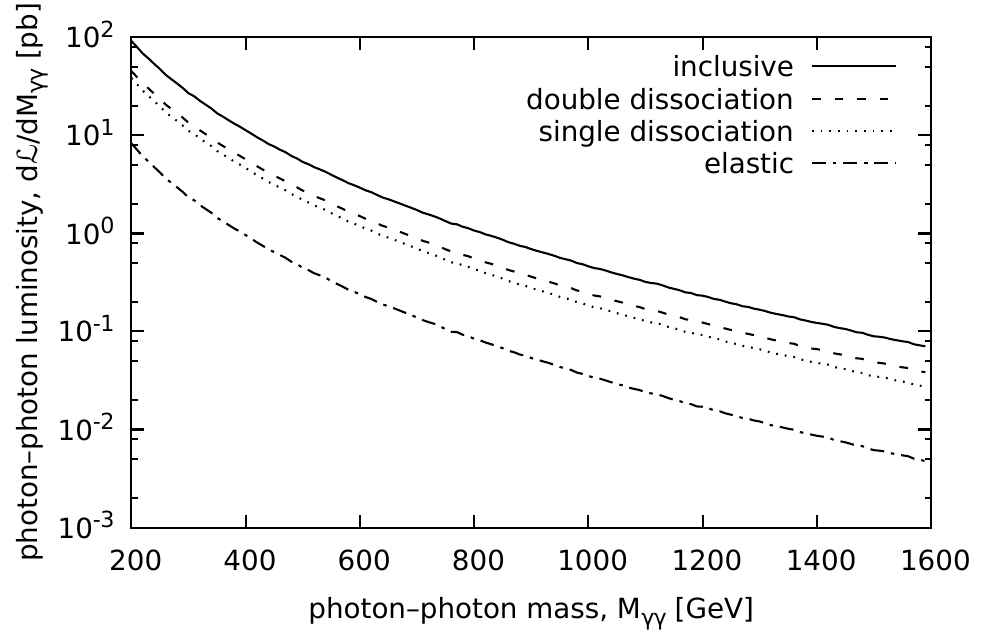}\\
  \caption{Photon--photon luminosity as a function of the mass of the photon--photon system for different production modes. 
  The inclusive process is the sum of all the other processes.}
  \label{fig:lumi_M}
\end{figure}

\begin{figure}[b]
  \includegraphics[width=\linewidth, page=2]{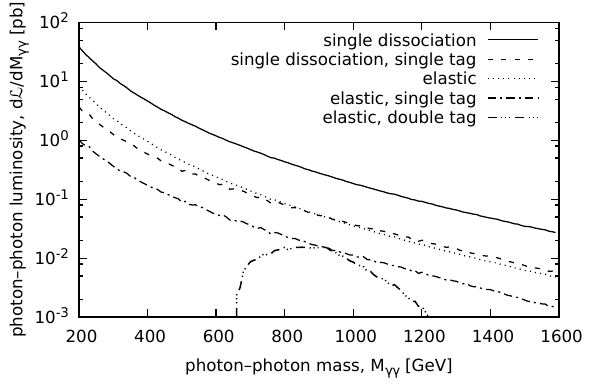}\\
  \caption{Photon--photon luminosity as a function of the mass of the photon--photon system for different measurement modes.
  }
  \label{fig:lumi_acc_M}
\end{figure}

Since in the two-photon process the photons must be emitted by both interacting protons,
one can distinguish three production modes:
\begin{itemize}
  \item fully elastic process: both photons are emitted leaving the protons intact,
  \item single dissociation: one photon is emitted elastically and the other one inelastically,
  \item double dissociation: both photons are emitted inelastically.
\end{itemize}

The photon--photon luminosities for different mechanism and experimental signatures are shown in Fig.~\ref{fig:lumi_M}.
For the mass of 750~GeV the total effective luminosity equals 1.33~pb,
where the double-dissociation, single-dissociation and elastic contributions are 0.69, 0.54 and 0.11~pb, respectively.
The single-dissociation contribution takes into account the factor of two because of the symmetry of the initial state
(each of the two protons can dissociate).

In addition, the elastically emitted protons may or may not be registered in the forward proton detectors.
This leads to three possible event signatures: 
\begin{itemize}
  \item double tag (exclusive measurement): protons are registered in the detectors on both sides,
  \item single tag (semi-exclusive measurement): a proton is registered in the detectors on one of the sides,
  \item no tag: no protons are registered.
\end{itemize}

Using an approximate acceptance of the forward proton detectors based solely on the energy of the proton (neglecting its transverse momentum) \cite{Trzebinski:2014vha},
it is possible to estimate how a requirement of a given proton-tag signature affects the photon--photon luminosity.
Here, it is assumed that the detectors can register protons that have lost between 5\% and 10\% of their energy.
This range is quite conservative and
it has been chosen to illustrate the advantage of the semi-exclusive measurement over the fully exclusive one.

Fig.~\ref{fig:lumi_acc_M} shows the effective luminosities for different measurement modes.
Obviously, the double-tag signature can be obtained only for the fully elastic process, while the single-tag signature can be obtained for the elastic and single dissociative ones.
One can see that the conservative acceptance to a large extent compromises the possibilities of a double-tag measurement at mass of 750 GeV.
On the other hand, the acceptance for the semi-exclusive measurement extends over the full mass range%
\footnote{%
In fact, an additional limitation is present because of the requirement to measure the final-state photons,
where the acceptance of the electromagnetic calorimeters plays a role.}.
In addition, the fact that the single dissociation production mode contributes to the semi-exclusive measurement greatly enhances the resulting photon--photon luminosity.

\section{Signal and background models}

In the following analysis it is assumed that the 750~GeV resonance is produced in two-photon processes, as described in the previous section.
The model contains three free parameters: the cross section for the $\GG\to\GG$ production, the mass and the width of the resonance.
The mass was chosen as 750~GeV and the width as 45~GeV, following \cite{Aaboud:2016tru}.
The value of the cross section does not affect the majority of the presented results and is chosen arbitrary, as discussed later on.
The transverse momenta of the virtual photons are neglected.
It is also assumed that the resonance decays only to photons and that the decay is isotropic.

For the fully exclusive measurement the dominant background was identified as photon pairs produced in hard non-diffractive events overlaid with independent pile-up events of soft single diffraction \cite{Fichet:2014uka}.
It is reasonable to expect the same situation also for the semi-exclusive measurement discussed here.
Therefore, the background events are simulated with Pythia 8.215 \cite{Sjostrand:2014zea} configured for $gg\to\GG$ and $q\bar q \to \GG$ parton level processes at $\sqrt{s}=13$~TeV.

The simulation of pile-up is performed for a given value of pile-up multiplicity $\mu$.
The actual number of soft diffractive protons in a given event is generated from a Poissonian distribution,
assuming the inelastic $pp$ cross section of 77~mb and the single diffractive cross section of 5.6~mb. 
The transverse momenta of the diffractive protons are neglected,
while the distribution of $\xi$,
the relative momentum loss of the proton,
is assumed to be of $1/\xi$ shape between $\xi_\text{min}=1.6\cdot10^{-9}$ and 1.
The $\xi_\text{min}$ value corresponds to the threshold for the $pp\to pp\pi^0$ process.
In order to make the analysis realistic, pile-up protons are added both to the background and to the signal events.

In the analysis the following experimental effects were taken into account:
\begin{itemize}
  \item acceptance of the forward proton detectors: an event is accepted if a proton with energy loss between 5\% and 10\% and $p_z>0$ was present (only detectors on one side of the interaction point are considered),
  \item forward detector energy resolution \cite{bib:afp},
  \item electromagnetic calorimeter energy resolution \cite{Aad:2008zzm},
  \item longitudinal spread of the vertex
    (affects the photon pseudo-rapidity reconstruction,
    since in two-photon events the vertex reconstruction may not be possible).
\end{itemize}

\section{Semi-exclusive measurement}

For an exclusive measurement a typical procedure is to use the information provided by the forward proton detectors to calculate the properties of the centrally produced state and to compare them with the direct measurement.
This approach is not possible for the proposed semi-exclusive measurement,
since the single-proton measurement is not sufficient to calculate the momentum of the central state.
However, the reversed procedure is possible: from the central state measurement one can estimate the energy of the forward proton, ${\xi_\GG}$,
which can than be compared with the measured value: $\xi_\text{p}$.
The relative energy of the forward proton can be calculated as:
\begin{equation}
  \nonumber  
  \xi^{\pm}_{\GG} 
  = (E_\GG \pm p_\GG)/{\sqrt{s}},
\end{equation}
where it is assumed that the photon--photon system has no transverse momentum.

\begin{figure}[b]
  \includegraphics[width=\linewidth, page=3]{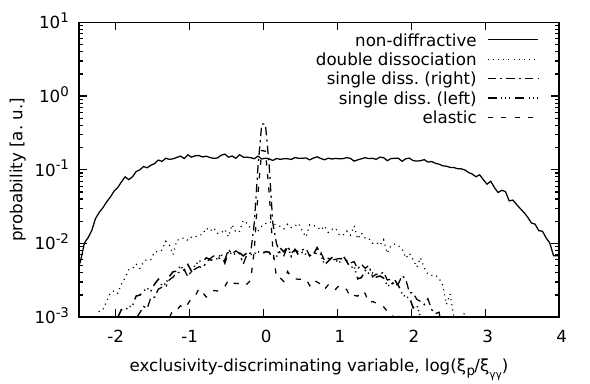}
  \caption{Distributions of the discriminating variable for different processes for $\mu=20$.}
  \label{fig:discrimination}
\end{figure}

It was checked that the signal events exhibit a strong correlation between $\xi_\GG$ and $\xi_\text{p}$,
while for the background the $\xi_\GG$ vs $\xi_\text{p}$ distribution is approximately flat.
The shape of the correlation suggests a choice of $\Delta=\log(\xi_\text{p} / \xi_{\GG})$ as the optimal discriminating variable.
The $\Delta$ distribution for the discussed processes, assuming the pile-up multiplicity of $\mu=20$, is shown in Fig.~\ref{fig:discrimination}.
For the elastic and single-dissociation signal events the distributions exhibit very strong peaks around zero.
On the other hand, flat distributions are observed for the non-diffractive production,
inelastic two-photon production and single-diffractive production with the intact proton on the wrong side.

\begin{table}[b]
  \caption{Background rejection factors for different values of the pile-up multiplicity.}
  \centering
  \begin{tabular}[]{rcccc}
    \toprule
    pile-up multiplicity & 5 & 10 & 20 & 40 \\
    background reduction factor & 260 & 130 & 66 & 34 \\
    \bottomrule
  \end{tabular}
  \label{tab:rejection}
\end{table}

Fig.~\ref{fig:discrimination} suggests the signal selection defined by $|\Delta| < \Delta_\text{cut}$,
with $\Delta_\text{cut} = 0.2$.
However, since the present analysis is relatively simple,
a more conservative value of $\Delta_\text{cut} = 0.5$ is taken,
which partially takes into account effects that are
not explicitly included in these considerations.
With this cut value, the signal efficiency is around 25\% and its dependence on the pile-up multiplicity is very small.
On the other hand, background rejection depends very strongly on the $\mu$-value, as shown in Tab.~\ref{tab:rejection}.

\begin{figure}[t]
  \includegraphics[width=\linewidth, page=2]{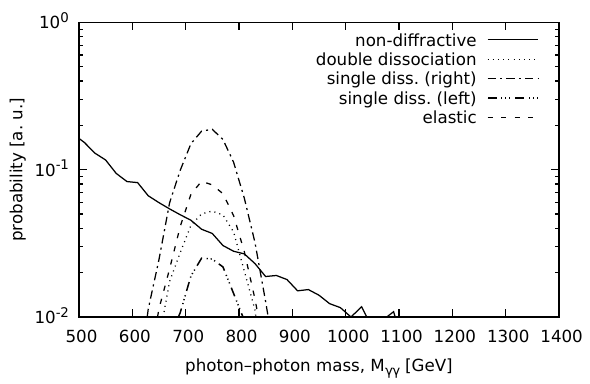}
  \caption{Photon--photon mass distribution for the signal and the background processes for pile-up multiplicity $\mu=20$ after the (semi-)exclusivity selection.}
  \label{fig:mass}
\end{figure}

In order to illustrate a possible result of the semi-exclusive measurement, 
Fig. \ref{fig:mass} shows the photon--photon mass distributions for the signal and background processes for the pile-up multiplicity of $\mu=20$.
Contrary to all previous results, here the absolute normalisation of the signal processes is needed.
It has been chosen so that the relative normalisation of the background and double-dissociation signal resembles the ATLAS results on the 750 GeV excess%
\footnote{The ATLAS analysis required a reconstructed vertex in the $\GG$ event, hence the possible contributions from elastic and single-diffraction processes should be suppressed.}.

One can observe that the proposed measurement method greatly improves the signal-to-background ratio.
In addition, the relative contributions of different two-photon production mechanisms in the final sample are different than before the selection (Fig.~\ref{fig:lumi_M}).
The inclusive production is dominated by double dissociation.
After applying the exclusivity selection the dominant contribution is due to single dissociation with the proton $p_z>0$.
The next contribution is the fully elastic one, then the fully inelastic one and finally the single dissociation with the proton $p_z<0$.

\section{Conclusions}

The above results show that a semi-exclusive measurement of the $\GG\to\GG$ processes,
in particular mediated via the hypothetical 750~GeV resonance, is a promising possibility.
The photon--photon luminosity is significantly higher in the single-dissociation mode than in the fully elastic one,
which will translate into higher statistics of the collected events.
In addition, the acceptance of the forward proton detectors is not as crucial as in the case of the double-tag measurements.

The proposed (semi-)exclusive selection provides a significant reduction of the background.
The improvement of the signal-to-background ratio strongly depends on the pile-up multiplicity:
for $\mu=5$ it is a factor of 65, while for $\mu=40$ it is around 8. 

In the presented analysis only the dominant background was considered and the source of the pile-up protons was treated quite roughly.
A future, more detailed analysis should also take into account additional background processes:
(semi-)exclusive photon--photon production via quark and $W$ boson loops, which is predicted by the Standard Model, diffractive photon--photon production \cite{Mariotto:2013kca} and backgrounds due to the photon misidentification.
Taking into account all these processes, and possibly a more sophisticated simulation of the experimental effects,
could allow more stringent exclusivity selection and thus a better background rejection and signal-to-background ration optimisation.

A proper treatment of the pile-up effects can have a two-fold consequence.
First, for the proton energy loss values of the order of 10\% processes other than single diffraction can contribute to pile-up effects.
Therefore, the background contribution will be increased.
On the other hand, taking into account the transverse momenta of the protons and the appropriate resolutions of the forward detectors will provide an additional constraint for the two-photon events selection.

This work was supported in part by Polish National Science Centre grant 2015/19/B/ST2/00989.

\newcommand{\reftitle}[1]{} 


\begin{thebibliography}{10}

    \bibitem{Aaboud:2016tru}
    ATLAS Collaboration,
    \reftitle{\textit{``Search for resonances in diphoton events at $\sqrt{s}$=13 TeV with the ATLAS detector''},}
    arXiv:1606.03833 [hep-ex].


    \bibitem{CMS:2015dxe} 
    CMS Collaboration [CMS Collaboration],
    \reftitle{\textit{``Search for new physics in high mass diphoton events in proton-proton
    collisions at 13TeV''},}
    CMS-PAS-EXO-15-004.

    \bibitem{Franceschini:2015kwy} 
    R.~Franceschini {\it et al.},
    \reftitle{\textit{``What is the $\gamma \gamma$ resonance at 750 GeV?''},}
    JHEP {\bf 1603}, 144 (2016).

    \bibitem{Ellis:2015oso} 
    J.~Ellis, S.~A.~R.~Ellis, J.~Quevillon, V.~Sanz and T.~You,
    \reftitle{\textit{``On the Interpretation of a Possible $\sim 750$ GeV Particle Decaying into $\gamma \gamma$''},}
    JHEP {\bf 1603}, 176 (2016).


    \bibitem{Lebiedowicz:2016lmn} 
    P.~Lebiedowicz, M.~{\L}uszczak, R.~Pasechnik and A.~Szczurek,
    arXiv:1604.02037 [hep-ph].

    \bibitem{bib:afp}
    ATLAS Collaboration,
    \reftitle{\emph{Technical Design Report for the ATLAS Forward Proton Detector}.}
    CERN-LHCC-2015-009.


    \bibitem{bib:pps}
    CMS-TOTEM Collaboration,
    \reftitle{\textit{``CMS-TOTEM Precision Proton Spectrometer''},}
    CERN-LHCC-2014-021.

    \bibitem{Khoze:2001xm} 
    V.~A.~Khoze, A.~D.~Martin and M.~G.~Ryskin,
    \reftitle{\textit{``Prospects for new physics observations in diffractive processes at the LHC and Tevatron''},}
    Eur.\ Phys.\ J.\ C {\bf 23}, 311 (2002).

    \bibitem{Fichet:2014uka}
    S.~Fichet, G.~von Gersdorff, B.~Lenzi, C.~Royon and M.~Saimpert,
    \reftitle{\textit{``Light-by-light scattering with intact protons at the LHC: from Standard Model to New Physics''},}
    JHEP {\bf 1502} (2015) 165.

    \bibitem{Fichet:2016pvq} 
    S.~Fichet, G.~von Gersdorff and C.~Royon,
    \reftitle{\textit{``Measuring the Diphoton Coupling of a 750 GeV Resonance''},}
    Phys.\ Rev.\ Lett.\  {\bf 116}, no. 23, 231801 (2016).

    \bibitem{Harland-Lang:2016qjy}
    L.~A.~Harland-Lang, V.~A.~Khoze and M.~G.~Ryskin,
    \reftitle{\textit{``The production of a diphoton resonance via photon-photon fusion''},}
    JHEP {\bf 1603} (2016) 182.


    \bibitem{Drees:1988pp} 
    M.~Drees and D.~Zeppenfeld,
    \reftitle{\textbf{``Production of Supersymmetric Particles in Elastic $e p$ Collisions''},}
    Phys.\ Rev.\ D {\bf 39}, 2536 (1989).

    \bibitem{Martin:2004dh} 
    A.~D.~Martin, R.~G.~Roberts, W.~J.~Stirling and R.~S.~Thorne,
    \reftitle{\textit{``Parton distributions incorporating QED contributions''},}
    Eur.\ Phys.\ J.\ C {\bf 39}, 155 (2005).

    \bibitem{Buckley:2014ana} 
    A.~Buckley, J.~Ferrando, S.~Lloyd, K.~Nordström, B.~Page, M.~Rüfenacht, M.~Schönherr and G.~Watt,
    \reftitle{\textit{``LHAPDF6: parton density access in the LHC precision era''},}
    Eur.\ Phys.\ J.\ C {\bf 75}, 132 (2015).

    \bibitem{Trzebinski:2014vha} 
    M.~Trzebi\'nski,
    Proc.\ SPIE Int.\ Soc.\ Opt.\ Eng.\  {\bf 9290}, 929026 (2014).

    \bibitem{Sjostrand:2014zea} 
    T.~Sjöstrand {\it et al.},
    Comput.\ Phys.\ Commun.\  {\bf 191}, 159 (2015).

    \bibitem{Aad:2008zzm} 
    ATLAS Collaboration,
    JINST {\bf 3}, S08003 (2008).

    \bibitem{Mariotto:2013kca} 
    C.~Brenner Mariotto and V.~P.~Goncalves,
    \reftitle{\textit{``Diffractive photon production at the LHC''},}
    Phys.\ Rev.\ D {\bf 88}, no. 7, 074023 (2013).

\end{thebibliography}
\end{document}